\documentclass[%
 reprint,https://www.overleaf.com/project/5cdedf2b9bd1c268f26e8fec
 amsmath,amssymb,
 aps,
]{revtex4-1}

\usepackage{graphicx}
\usepackage{dcolumn}
\usepackage{bm}
\usepackage{textgreek}
\usepackage{epstopdf}

\begin{document}

\preprint{APS/123-QED}

\title{Theoretical and experimental electron diffraction intensity maps for single crystal silicon from an ultrafast source}

\author{L.E. Malin}
\email{lemalin@asu.edu}
\author{W.S. Graves}%
\author{M. Holl}%
\author{J.C.H. Spence}%
 \affiliation{%
 Department of Physics, Arizona State University, Tempe, Arizona 85287, USA
}%
\author{E.A. Nanni}%
\author{R.K. Li}%
\author{X. Shen}%
\author{S. Weathersby}%
 \affiliation{%
	SLAC National Accelerator Laboratory, Menlo Park, CA 94025 USA
}%




\date{May 20, 2019}

\begin{abstract}
Electron diffraction through a thin patterned silicon membrane can be used to create complex spatial modulations in electron distributions. By precisely varying parameters such as crystallographic orientation and wafer thickness, the intensity of reflections in the diffraction plane can be controlled and by placing an aperture to block all but one spot, we can form an image with different parts of the patterned membrane, as is done for bright-field imaging in microscopy. The patterned electron beams can then be used to control phase and amplitude of subsequent x-ray emission, enabling novel coherent x-ray methods.  The electrons themselves can also be used for femtosecond time resolved diffraction and microscopy.  As a first step toward patterned beams, we demonstrate experimentally and through simulation the ability to accurately predict and control diffraction spot intensities.  We simulate MeV transmission electron diffraction patterns using the multislice method for various crystallographic orientations of a single crystal Si(100) membrane near beam normal. The resulting intensity maps of the Bragg reflections are compared to experimental results obtained at the Accelerator Structure Test Area Ultrafast Electron Diffraction (ASTA UED) facility at SLAC. Furthermore, the fraction of inelastic and elastic scattering of the initial charge is estimated along with the absorption of the membrane to determine the contrast that would be seen in a patterned version of the Si(100) membrane.

\end{abstract}

\pacs{Valid PACS appear here}
\maketitle


\section{\label{sec:intro}Introduction}

We describe the first experimental steps toward nanopatterned electron beam production that is expected to enable fully coherent ultrashort pulse x-ray generation from compact accelerators.  We accurately predict the diffraction efficiencies of electrons into particular Bragg spots through thin single-crystal silicon membranes using a multislice propagation method, and verify these predictions through experiments with femtosecond electron bunches produced by a photoinjector.  These experiments and simulations demonstrate that electron density can be spatially modulated at fine scales by controlling the diffraction parameters.  After briefly introducing the method and its impact on x-ray emission, we present calculations of dynamical diffraction of relativistic electrons, then compare the calculations with experimental results, and finally analyze inelastic effects.

The world's brightest x-ray sources are x-ray free electron lasers (XFELs) and synchrotron rings that use bunches containing up to a billion electrons each to produce powerful x-ray pulses.  Within a bunch, the electrons do not exhibit order at the x-ray wavelength scale so that the radiation each electron emits is out of phase with its neighbors.  The total output radiation consists of many wavetrains superimposed with random phase giving classic shot noise characteristics of intensity and spectral fluctuations.  Synchrotron undulator radiation retains the random phasing and resulting shot noise properties \cite{Schwinger1945,Schwinger1949} in the output radiation.  The spectrum only appears smooth when integrated over many shots.  XFELs rely either on self amplification of spontaneous emission (SASE\cite{Pellegrini}) which is amplification of the initial shot noise or, if available, an external coherent radiation seed pulse stronger than the shot noise that pushes the electrons to bunch at the desired photon wavelength.  As the electrons in an XFEL arrange themselves into a coherent density modulation in response to the growing radiation field, their emission phases begin to align, with a resulting x-ray intensity increase over synchrotron emission comparable to the number of electrons, or 6 to 9 orders of magnitude.  XFELS are thus very powerful and have opened many new scientific capabilities.  However, because no coherent seed source exists, XFEL radiation has well-known limitations including large fluctuations in time and spectrum due to the amplified shot noise of SASE, or in the case of self-seeding \cite{Geloni2010a,Geloni2010b}, fluctuations in seed power that is itself based on SASE as well as a complicated beamline setup.  

We are pursuing an alternative method of generating bright coherent x-ray beams through the production of electron beams that are bunched at x-ray wavelength scale prior to emitting radiation.  This allows fine control of the x-ray phase fronts potentially including frequency-chirped output, multiple colors, attosecond delays, and tunable pulse length and spectral width.  We have investigated two methods to modulate the beam, first examining arrays of nanoemitters at the cathode
\cite{Graves2012} and more recently exploring diffraction of the electron beam through thin lithographically etched crystals\cite{Nanni2018}. Both of these methods generate a transverse variation in electron density that must be converted to a longitudinal (i.e. time) variation to generate coherent radiation.  This conversion can be achieved through an emittance exchange (EEX) beamline.  We have previously analyzed this process including aberration correction optics to facilitate the exchange for emittance differences of up to four orders of magnitude\cite{Nanni2015}. Eventually we expect to use this method to convert the compact x-ray light source (CXLS) at Arizona State University to a coherent x-ray source. In the initial experiments reported here we determine the degree to which an electron beam can be controlled via diffraction as well as the quality and characteristics of the output electron beam.   

The purpose of the electron diffraction is to deliberately deflect portions of the electron beam off-axis and then, after some drift distance, block the unwanted parts. We call this method a dynamical beam stop. If just two strong Bragg beams can be excited, the Pendell\"{o}sung effect in multiple scattering \cite{Zuo2017} causes the two to exchange energy periodically as a function of sample thickness, i.e. as we vary the sample thickness, the direct beam becomes more intense as the diffracted beam dwindles and vice versa. This periodicity is determined by the extinction distance, which depends on the material, the scattering angle of the diffracted beam, and the electron energy. Knowing this, it is then possible to choose particular thicknesses of silicon (at multiples of the  extinction distance) at which practically all the energy has been transferred to either the Bragg diffracted or direct beam. A spatial pattern will then be imposed across the electron beam by using strips of silicon of the correct thickness running across the beam, separated by very thin non-diffracting almost transparent material. With the thickness of the strips chosen to diffract almost all energy into the Bragg diffracted beam, which then scatters outside the hole in a central aperture and so cannot contribute to image-formation, an image of the silicon formed using the direct beam only will appear dark within the strips, and bright between them, as for a bright-field image in microscopy. As will be shown, there is an advantage to using the direct beam in the final spatial modulation.

Depending on the parameters of the setup, it can be difficult to get a pure two-beam case, so the above extinction lengths serve as starting point; as beam energy can vary and the thickness of the blocking portion is more or less fixed, to get a majority of the incident beam into the intended reflection requires fine tuning the sample's crystallographic orientation. In order to maximize the contrast of the spatial pattern, one must consider more than the kinematic regime for electron diffraction; in the dynamic regime we encounter, a significant number of diffraction spots can be excited by the incident beam. This is particularly true at the relatively high energies (for electron diffraction) of a photoinjector, diminishing the applicability of the two-beam analytic theory and bolstering the need for simulations to determine optimal crystallographic orientation for a given membrane thickness and beam energy. A dynamical multiple beam approach is needed, which is described next.

\section{\label{sec:MSM}Multislice Method Theory}

    The mathematical tools to calculate the intensities of diffracted beams have been around for nearly 100 years (a brief overview can be found in Humphreys\cite{Humphreys1979}). In particular, the multislice method, proposed over half a century ago\cite{CowleyandMoodie1957}, has been used to simulate experimental diffraction patterns to much success\cite{Zuo2017}. The scattering across the depth of the crystal can then be calculated, including multiple scattering events of single electrons.

    We apply the multislice method by using Sch\"{o}dinger's equation iteratively and dividing the crystal's potential into multiple layers along the electron's direction of travel.  The electron wavefunction $\psi_n(x,y)$ can be calculated at the exit of the crystal, representing the exit beam downstream of the crystal. For the $n^{th}$ layer, the wavefunction is
    \begin{equation}\label{eq:multislice}
    \psi_{n+1}(x,y)=p_n(x,y) \ast [t_n(x,y)\cdot \psi_n(x,y)] 
    \end{equation}
    where $p_n(x,y)$ is the Fresnel propagator, $t_n(x,y)$ is the transmission function, and $\ast$ is the 2-dimensional convolution. In the physical optics interpretation \cite{CowleyandMoodie1957}, the propagator accounts for near-field diffraction while the transmission function describes a phase grating.
    
    A less computationally intensive form of Eq. (\ref{eq:multislice}) can be had by applying the Fast Fourier Transform in conjunction with the convolution operator theorem\cite{Ishizuka1977}. The electron wavefunction is then given by
    \begin{equation}\label{eq:fftmultislice}
    \psi_{n+1}(x,y)=\mathcal{F}^{-1}\{P_n(k_x,k_y) \cdot \mathcal{F}[t_n(x,y)\cdot \psi_n(x,y)]\} 
    \end{equation}
    where $P_n(k_x,k_y)$ is the Fourier transform of the real-space propagator, while  $\mathcal{F}$ and $\mathcal{F}^{-1}$ are the Fourier and inverse Fourier transforms respectively. The real-space propagator (for small crystal tilt angles $\approx 1^\circ$) is
    \begin{equation}\label{eq:prop}
    P_n(k_x,k_y)=\exp[-i \pi \lambda \Delta z +
        2 \pi i \Delta z \alpha(k_x,k_y,\theta_x,\theta_y)]
    \end{equation}
    where $\alpha(k_x,k_y,\theta_x,\theta_y)=k_x \tan \theta_x + k_y \tan \theta_y$, $k_x$ and $k_y$ are the $x$ and $y$ components of the wavenumber, $\theta_x$ and $\theta_y$ are the $x$ and $y$ components of the sample tilt, and $\Delta z$ is the slice thickness. Further, the transmission function is 
    \begin{equation}\label{eq:trans}
    t(x,y,\Delta z)= \exp[i\sigma V(x,y) \Delta z] 
    \end{equation}
    where $\sigma$ is the relativistic electron interaction constant given by $\sigma =\frac{ 2\gamma m_0|e|\lambda}{4\pi\hbar^2}$, with $m_0$ the electron rest mass, $\gamma$ the Lorentz factor, $e$ the electron charge, $\lambda$ the relativistic electron wavelength, and $\hbar$ the reduced Planck constant. 
    
    In Eq. (\ref{eq:trans}), $V(x,y)$ is the crystal potential projected along the beam direction z that describes the potential within a distance $\Delta z$ of the current layer; we approximate this as a sum of all the individual atomic potentials in the layer, which can be treated as Fourier coefficients. These are proportional to the the electron scattering factor and are weighted by a Debye-Waller temperature factor $B$ according the expression $\exp[-Bs^2]$, where $s = \frac{sin \theta}{\lambda}$, $\theta$ is the real-space scattering angle associated with the reciprocal space coordinate $s$, and $\lambda$ is the relativistic electron wavelength. The temperature factor causes increased attenuation of high-angle scattering with increasing temperature. There exist many parameterizations that allow the calculation of the electron scattering factor \cite{Kirkland2010} and the Debye-Waller factor\cite{DWfactor1991} for various elements and temperatures.

    Partial coherence occurs when there is a spread in the momentum of the incident electron beam and is related to the concept of emittance. In the absence of a magnetic field, we can write $p_x = m_0c\beta\gamma x'$ where $m_0$ is the electron rest mass and $c$ is the speed of light, $x' = dx/dz$ (same relation holds for $y$), as well as $\vec p = \hbar \vec k$. In this context the normalized emittance of the beam is given by 
    
    \begin{equation}\label{eq:normEmit}
    \sigma_n^x= \frac{\hbar}{m_0c} \sqrt{\langle{x^2}\rangle \langle{k_x^2}\rangle - \langle{xk_x}\rangle^2}
    \end{equation}
    where $\alpha$ and $\gamma$ are the relativistic factors and $x$ is the particle position and $x'$ is the angle of the trajectory. A similar expression holds for the y-direction. 
    
    To match the momentum spread present in physical beams, we first consider an electron plane wave with wavefunction $\psi(\vec x, \vec {k_i}) = \exp[2\pi \vec {k_i} \cdot \vec x]$ where $\vec {k_i}$, is the angular deflection of the incoming beam from normal. To include this partial coherence, we sum over the angles and apply a weighting function $p(\vec {k_i})$; in this case, the beam can be approximated with a Gaussian weighting. The resulting intensity is

    \begin{equation}\label{eq:parco}
    I(\vec x)=\frac{1}{N} \sum_{i} p(\vec {k_i})|\psi_t(\vec x, \vec {k_i})|^2
    \end{equation}
    where $\psi_t(\vec x, \vec {k_i})$ is the transmitted wavefunction at exit from the crystal and ${N}$ is the number of angles included in the sum.
    
    We have implemented this multislice algorithm in MATLAB and benchmarked it against a version of the highly developed multislice JEMS code modified by its author to track relativistic electrons\cite{JEMS}.  The codes are in good agreement for the range of parameters we are interested in.  Our next step is to determine the experimental beam and crystal parameters that need to be simulated.

\section{\label{sec:UED}UED Experiment}

    Data was collected at the SLAC Accelerator Structure Test Area Ultrafast Electron Diffraction (ASTA UED) facility, which uses an RF photoinjector (gun) with a focusing solenoid magnet as the electron source\cite{ASTAUED}. The gun is capable of RMS pulse lengths on the scale of 100 fs at few MeV energies with an energy spread of $7.5 \times 10^{-4}$ and a repetition rate of 180 Hz. The minimum achievable spot size at the sample is approximately 5 \textmu m. For this experiment, a kinetic energy of 2.26 MeV was used.  An adjustable collimator is located 0.56 m from the cathode;  we used a 90.7 \textmu m aperture, giving a charge of 10 fC per shot, though other larger aperture and higher charge combinations were examined. Using a solenoid scan with the second solenoid located at 1.0 m from the cathode, the normalized emittances were calculated to be 1.1 nm-rad for $\epsilon_n^x$ and 5.4 nm-rad for $\epsilon_n^y$. This second solenoid served to focus the beam onto the 6-axis sample holder located at approximately 1.36 m.  The sample holder is capable of $\pm 30^{\circ}$ rotation along the x and y axes (pitch and yaw) as well as $\pm 1^{\circ}$ rotation along z axis. A Norcada UberFlat single crystal Si(100) membrane with a uniform thickness of just 200 nm and size of 100 \textmu m $\times$ 100\ \textmu m was inserted in the holder. To minimize the post-processing required to determine the orientation of the pitch and yaw axes of the sample holder relative to the crystal plane of the membrane, great effort was made in mounting the sample to the holder with less than $1^\circ$ of roll - a value that could be corrected in situ.

\begin{figure}
  \centering
  \includegraphics[width=\linewidth]{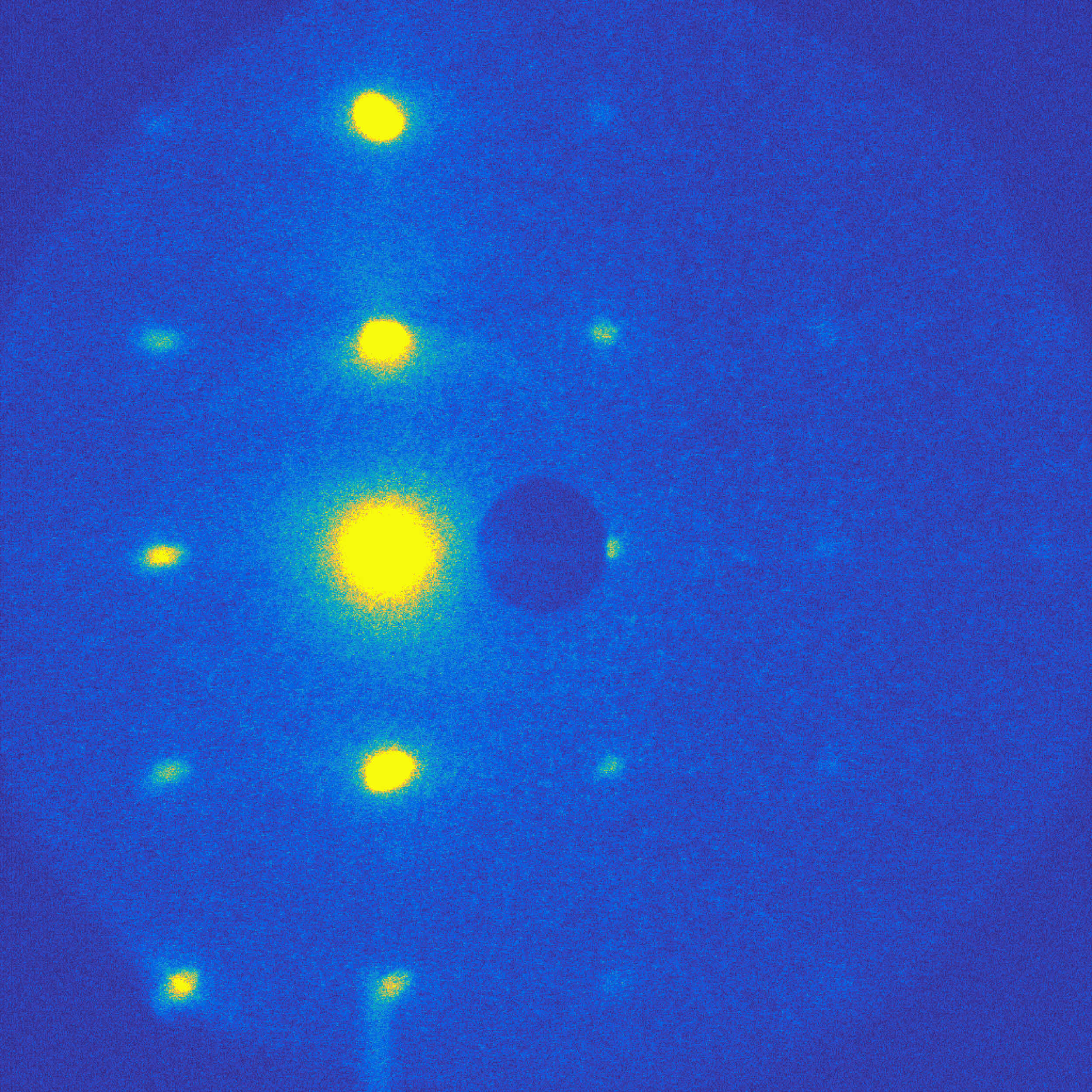}
    \caption{Transmission electron diffraction pattern at 2.26 MeV from Si(100) (contrast adjusted). This image was taken with a gain of 75 and an exposure of 0.102 s, and thus is representative of multiple shots of the ultrafast source running at 180 Hz. There are 20 Bragg reflections in the diffraction pattern; going left to right and top to bottom, the following reflections are of interest: row 1 column 1 $(\overline{6}20)$; row 2 column 2 $(\overline{2}20)$; row 3 column 2 $(000)$; row 4 column 2 $(2\overline{2}0)$; and row 5 column 1 $(2\overline{6}0)$.  The dark circle is a hole in the scintillator.}
    \label{fig:diffPattern}
\end{figure} 

    Beyond the holder, there is a 3.12 m drift to a YAG screen, which is imaged using a Andor iXon Ultra 888 EMCCD. The pixel size was calculated to be 36 \textmu m in real space (using a 4.5 mm aperture in the scintillator as a reference) and 0.00179 $\AA^{-1}$ in reciprocal space (using the spacing between the (440) and $(4 \overline{4}0)$ spots). Fitting the beam without the sample, the RMS reciprocal space width $\sigma_k$ was 0.0133 $\AA^{-1}$, which corresponds to an RMS angle $\sigma_{x^\prime}$ of 61 \textmu rad at the sample.   

    \begin{figure*}[!htb]
    \centering
    \includegraphics[width=\textwidth]{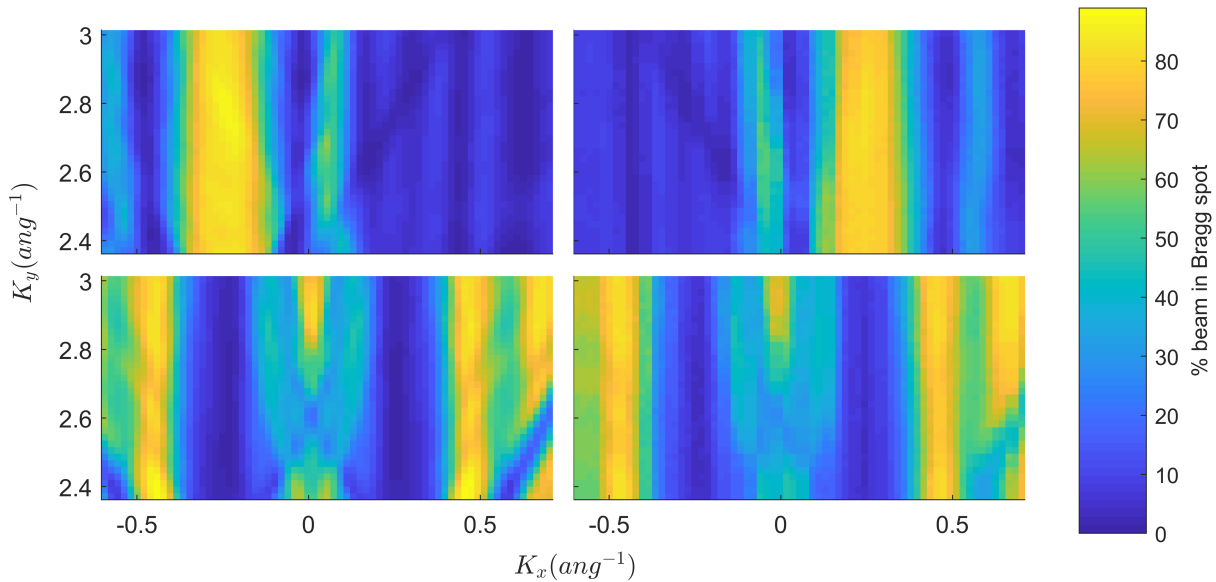}
    \caption{Simulated (left) and experimental (right) intensity maps for two of the Bragg reflections: $(2\overline{2}0)$ above and $(000)$ below scanned over a range of sample pitch and yaw angles. The percentages in the color bar refer to the fraction of the elastic scattering. }
    \label{fig:IntensityMaps}
    \end{figure*} 
    
    It is important to know the angle between the normal of the sample crystal axes and the beam direction in order to match simulations to experiment. Thus, it was necessary to determine the holder angle settings that corresponded to beam normal; this was determined by first tilting the sample until the diffraction pattern had symmetric intensities, setting the beam direction near the $[001]$ zone axis. After this, alignment intensity maps were produced by scanning both tilt angles, pitch and yaw, for a small set of tilts surrounding the proposed normal. Symmetry in the intensity map was used to determine the beam normal. The sample was then tilted through a wide range of pitch and yaw settings that matched the collection of wavenumber ($k_x$,$k_y$) values of interest while recording the Bragg spot intensities on the YAG scintillator. One Bragg pattern is shown in Fig. \ref{fig:diffPattern} with contrast enhanced to show all spots and, importantly, the diffuse inelastic scattering between spots.
    
    We acquired many such diffraction patterns or rocking curves, recording the Bragg intensity as a function of pitch and yaw of the sample. We then used the data to produce intensity maps for particular Bragg spots as a function of 2D tilts, and compare these intensity maps to our multislice simulations. The CCD background and dark current were subtracted off by taking the average of multiple background images with the UV laser off and the RF to pulse. Furthermore, an averaged inelastic background was subtracted from each diffraction image to get the elastic contribution.  The centers of all visible diffraction peaks were set at the local maxima and all counts within 25 pixels (or slightly greater than $3\sigma$) of the center were summed over for the respective reflection. These were then normalized by using the total counts in all the visible diffraction peaks. Through these measurements, our goal is to find particular angles where, for a given sample thickness (200 nm here), the intensity into one of the low order spots is maximized and the forward scattered beam (000) spot in the same direction as the incident beam is minimized: the so called "two beam" condition.

\section{\label{sec:Results}Results and Discussion}

    In this section we discuss the intensity maps from experiment and compare them to multislice simulations modeling diffraction over the same sets of angles and with the same beam size and emittance.  We are also able to separate the elastically scattered component of the original beam in the experimental data from the inelastic component and make an estimate for the absorption of the 2.26 MeV beam in the 200nm Si(100) membrane. 

\subsection{Comparison of Experiment to Simulations}

    \begin{figure*}[!htb]
	\centering
	\includegraphics[width=\textwidth]{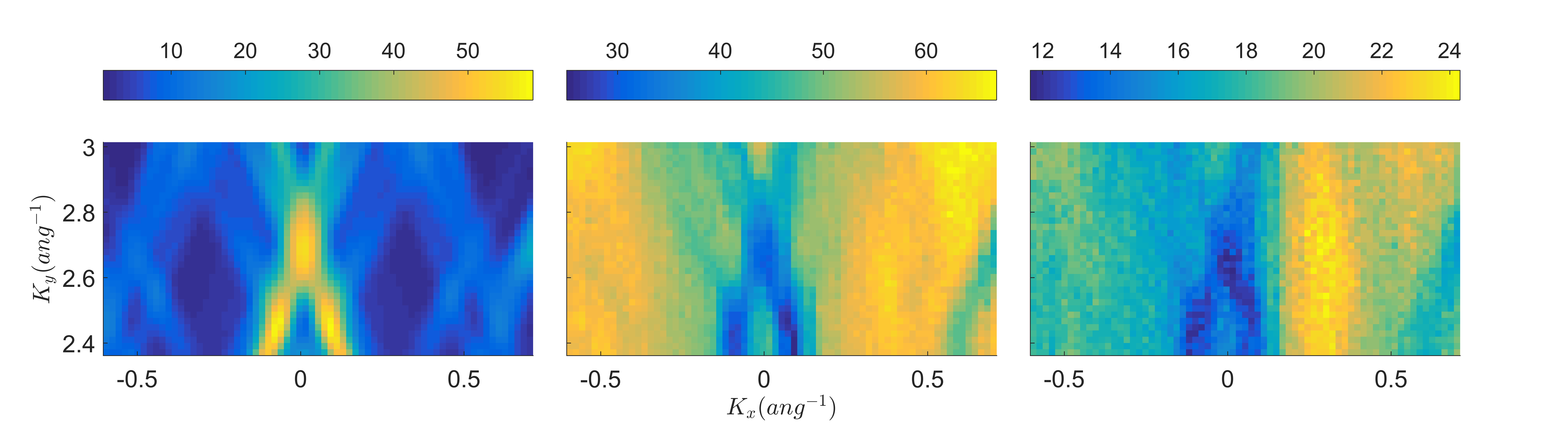}
	\caption{Intensity maps showing percentage of beam scattered in 3 different ways: (left) simulation of elastic scattering intensity into all higher order Bragg reflections that are not included in the 20 spots we can measure; (middle) experimental elastic intensity measured in those 20 reflections we can see (percentage of incident charge); (right) the experimental inelastic scattering measured in the image  as a percentage of the initial charge.}
	\label{fig:elasticVSinelastic}
\end{figure*} 

    The intensity maps are shown in  Fig. \ref{fig:IntensityMaps} for both simulation (left) and experiment (right). There are no free parameters in these simulations, which use the experimental parameters to propagate electrons through the simulated membranes in slices 5.43 nm thick (the same as the lattice parameters of Si(100)). The agreement between simulation and experiment is excellent. This gives confidence that we understand a number of important parameters and effects including the properties of the Si membranes, the diffraction physics, and the beam properties from the photoinjector. It gives us confidence that we can accurately predict performance of similar lithographically etched membranes that will be used to nanopattern the electron beam. The extinction of the $(000)$ beam shown near $k_x = 0.3 \ \AA^{-1}$ means that very strong contrast between different thicknesses of membrane can be generated, which is the desired condition for strong nanobunching.
    
        The maxima of the simulated Bragg intensities are slightly larger for both Bragg spots modeled, with the simulated $(000)$ reflection having a maximum of 89\% and the experimental reflection 83\%. The simulated maps also show more detail than the experimental patterns; for instance, the two valleys that start near the bottom of $k_x = 0.5 \ \AA^{-1} \textnormal{ and } k_x = -0.5 \ \AA^{-1}$ are less pronounced in the experimental data. This can be explained by the clipping of the $(\overline{6}20)$ and $(2\overline{6}0)$ reflections by the lens (see Fig. \ref{fig:diffPattern}) of the camera. Each reflection has its respective value where it receives a majority of the intensity; clipping causes the contribution to the $(000)$ reflection to appear inflated and the valleys shallow. Of note is the elastically scattered minimum of 3.5\% for the $(000)$ beam, which in the context of a patterned membrane would allow for significant contrast. It will become apparent in the next section why we should seek contrast in the $(000)$ beam versus one of the higher order reflections.
    
    To get the necessary reciprocal space resolution, 8-by-8 unit cells of Si(100) were used, while to get the real space resolution to properly simulate the projected potential of the increased number of unit cells required a $512 \times 512$ grid. The simulated maps were also normalized to the 20 Bragg reflections in the field of view of the diffraction pattern (for example see Fig. \ref{fig:diffPattern}). Note that the $k_x$ and $k_y$ axes are rotated 135 degrees relative to the $(100)$ plane of the crystal as the crystal planes of the membrane were rotated relative to the pitch and yaw planes of the holder.

\subsection{Inelastic vs Elastic Scattering}

    During the analysis of the experimental data we determined that a significant amount of charge incident on the membrane is not contained within the Bragg spots, and is either absorbed, elastically scattered into high-order spots outside our field of view, or inelastically scattered.  Our multislice simulation can address elastic scattering into high order spots for different tilt angles to account for that portion of the charge.  Then by comparing that quantity with the experimental data, we can deduce the fraction that is inelastically scattered and absorbed. To get the elastically scattered portion of the experimental pattern, we summed the counts from all the diffraction spots after the previously mentioned background removal was performed. The full charge of each image was calculated by removing the background due to the CCD and dark current and then summing the counts.
    
\begin{figure}
  \centering
  \includegraphics[width=\linewidth]{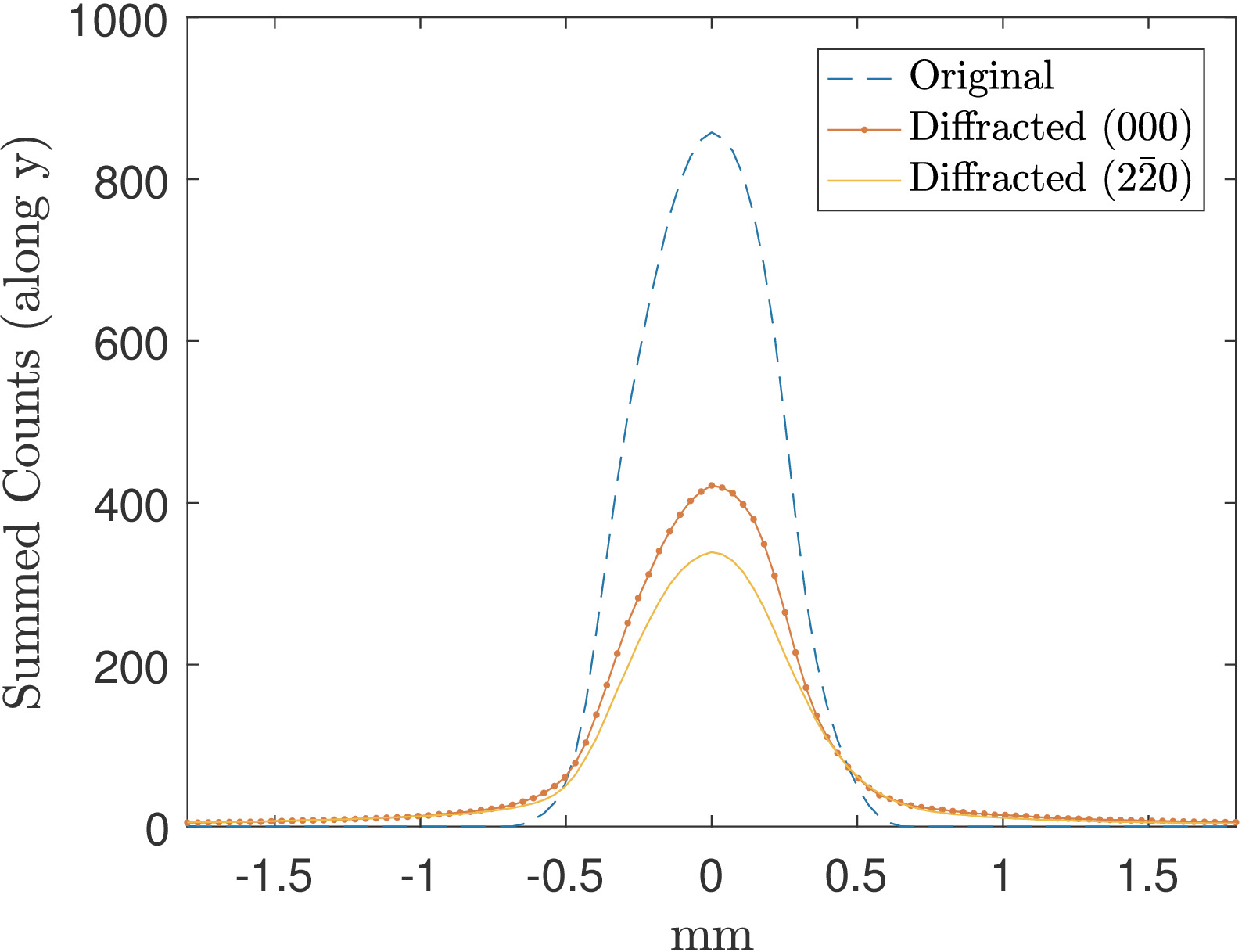}
    \caption{A comparison of the summed profile of the electron beam before diffraction to after diffraction. The long tails of the diffracted beams indicate inelastic electron-plasmon scattering.  Note that the beams taken after diffraction are the direct $(000)$ beam and the $(2\overline{2}0)$ beam, which were taken from two separate diffraction patterns. The local background was removed for all cases. The image of the original beam was taken with a gain of 75 and exposure of 0.102 s; the image of diffraction patterns was taken with a gain of 150 and an exposure of 0.102 s, though the gain differences have been compensated.}
    \label{fig:profiles}
\end{figure}

    Looking at the middle plot of Fig. \ref{fig:elasticVSinelastic}, we can see that a maximum of 67\% of the initial electron pulse is elastically scattered. From the left plot of the same figure, we can see that the variation in the measured intensity of the middle plot can be attributed to elastic scattering into higher-order reflections. Examining the total charge in each diffraction image and subtracting the elastic component, we calculate the percentage of inelastic scattering of the initial beam (right plot of Fig. \ref{fig:elasticVSinelastic}). We see a similar pattern to the elastic scattering (middle plot), with a valley in the center where one would expect diffraction into higher-order reflections. From this we infer that the inelastic scattering is occurring near the excited spots outside our field of measurement, which is consistent with plasmon diffuse scattering. In this type of scattering, incident electrons excite plasmons from valence electrons in the material, causing the incident electron to lose energy and scatter through angles smaller than the Bragg angle. This type of scattering is isotropic about the Bragg peaks with a Lorentzian distribution and half angle of $\theta_E = \Delta E/2E$ where $\Delta E$ is the energy loss and $E$ is the beam kinetic energy\cite{Egerton2011}. This energy loss is equal to the plasmon energy, which for silicon is 16.7 eV\cite{Egerton2009}. At 2.26 MeV, this gives a half angle of 3.7 \textmu rad. Examining Fig. \ref{fig:profiles}, there is a tail as one would expect when comparing a Gaussian to something with the expected Lorentzian component. Figure \ref{fig:profiles} also confirms that the plasmon diffuse scattering shifts to the brightest reflection as both the diffracted $(000)$ and $(2\overline{2}0)$ reflections have similar tails.
    
    Returning to the discussion of a patterned dynamical beam block, to maintain the Guassian profile of the beam and minimize the contrast reduction, the length over which the electrons interact with the material must be minimized. From the tails in the diffracted profiles of Fig. \ref{fig:profiles} and the elastic and inelastic percentages of Fig. \ref{fig:elasticVSinelastic}, we see that even at 200 nm and the high energies we are working with, a significant portion of the incident beam is lost to inelastic processes. If we were to make a pattern in a membrane with this thickness or greater, the bleed over from inelastic scattering would reduce the contrast. The membrane must be as thin as possible while still maintaining the integrity of the pattern etched into it. At the same time, the thinner the material becomes, the less likely it is for the incident beam to be deflected away from the direct beam (000). For this reason, the blocking portions of the membrane should be thicker than the portions that will be eventually used to form the image. This would mean using the direct $(000)$ reflection in the final patterning, i.e. using bright field imaging. Furthermore, to limit the plasmon diffuse scattering from higher order reflections overlapping the imaging beam, the membrane should be oriented so that the excited reflections in the blocked portion are relatively far from the direct beam, though too large of a tilt could affect the sharpness and modulation of the image produced from the patterned membrane.
    
\subsection{Absorption}
    
    To estimate the absorption occurring in the crystal, we can add an imaginary term to the projected potential of the crystal. This gives rise to an exponentially decaying damping term on the intensity\cite{SpenceMD}: $exp[{-\frac{t}{\lambda}}]$ where the mean free path is given by
        
    \begin{equation}\label{eq:DampTerm}
    \lambda=\frac{(hc)^2k_{0_z}}{4\pi m_e c^2 e V_0^\prime}.
    \end{equation}
    In the above, $k_{0_z}$ is the z component of the wave vector, which for high energy electrons is approximately the electron wavelength $\lambda_e$. $V_0^\prime$ is the imaginary part of the potential and is in unit of Volts. In Radi\cite{Radi1970}, the imaginary portion of the atomic potential was calculated using Hartree-Fock-Slater (HFS) atomic functions for 100 keV. These can be scaled to the energy used in the experiment, yielding a mean free path of 3.57 $\mu m$, which gives an absorption of 5.5\% at 200nm. Together, the elastic and inelastically scattered components account for about 88\% of the original beam. The remaining approximate 12\% of the beam is being absorbed by the crystal, going into inelastic processes that are outside the measured area of the image, or were subtracted during the CCD background and dark current removal procedure.
    
\section{Conclusions}
    Simulations of ultrafast relativistic electron diffraction rocking curves using the multislice method show excellent agreement with experimental results, leading to confidence in predictions of diffracted beam intensity as a function of sample thickness.  These results also demonstrate that RF photoinjectors produce high quality electron beams meeting the requirements for nanopatterned beams.  The experimental intensity map for a 200 nm thick Si(100) membrane was measured and compared to the simulated map. The agreement between the elastic scattering of experiment and simulation was found to match - as would be expected from the product of a well-developed and long-lived theory. The fraction of elastic and inelastic scattering as a percentage of the charge in the electron packet was found to be respectively $~67\%$ and $~21\%$ with the remaining percentage being some combination of absorption and inelastic processes. From this, we were able to determine that the spatial modulation should be formed from the bright field image and that the thickness of the dynamical beam block used to form this image should be minimized. This image will then serve as input into an EEX beamline to provide the prebunching needed to shift the inverse-Compton scattering based CXLS under construction at ASU into the super-radiant regime as a compact x-ray free electron laser (CXFEL).
    
\section{Acknowledgements}
     We gratefully acknowledge financial support from NSF awards 1632780 and 1231306, the BioXFEL Science and Technology Center, and the SLAC MeV UED facility, which is supported in part by the DOE BES SUF Division Accelerator \& Detector R\&D program, the Linac Coherent Light Source (LCLS) Facility, and SLAC under contract nos. DE-AC02-05-CH11231 and DE-AC02-76SF00515.
    

%

\end{document}